\documentclass[12pt,preprint]{aastex}

\newcommand{\myemail}{shogo@optik.mtk.nao.ac.jp}
\newcommand\ionn[2]{#1$\;${\small\rmfamily{#2}}\relax}%

\begin{document}

\title{Herbig Ae/Be Stars in the Magellanic Bridge}

\author{Shogo Nishiyama\altaffilmark{1,2}, 
Yasuaki Haba\altaffilmark{2}, 
Daisuke Kato\altaffilmark{2},
Daisuke Baba\altaffilmark{2},
Hirofumi Hatano\altaffilmark{2},
Motohide Tamura\altaffilmark{1},
Yasushi Nakajima\altaffilmark{1},
Akika Ishihara\altaffilmark{1,3},
Tetsuya Nagata\altaffilmark{4},
Koji Sugitani\altaffilmark{5},
Noriyuki Matsunaga\altaffilmark{6}, 
Hinako Fukushi\altaffilmark{6}, 
Nobuhiko Kusakabe\altaffilmark{7}, 
and Shuji Sato\altaffilmark{2}
}

\altaffiltext{1}{National Astronomical Observatory of Japan, 
Mitaka, Tokyo 181-8588, Japan; \myemail}

\altaffiltext{2}{Department of Astrophysics, Nagoya University, 
Nagoya 464-8602, Japan}

\altaffiltext{3}{Department of Earth and Planetary Science, School of Science, 
University of Tokyo, Bunkyo-ku, Tokyo 113-0033, Japan}

\altaffiltext{4}{Department of Astronomy, Kyoto University, 
Kyoto 606-8502, Japan}

\altaffiltext{5}{Graduate School of Natural Sciences, Nagoya City University,
Nagoya 467-8501, Japan}

\altaffiltext{6}{Institute of Astronomy, School of Science, 
University of Tokyo, Mitaka, Tokyo 181-0015, Japan}

\altaffiltext{7}{Department of Astronomical Sciences,
Graduate University for Advanced Studies (Sokendai),
Mitaka, Tokyo 181-8588, Japan}


\begin{abstract}

We have found Herbig Ae/Be star candidates
in the western region of the Magellanic Bridge.
Using the near infrared camera SIRIUS 
and the 1.4 m telescope IRSF,
we surveyed $\sim 3\fdg0 \times 1\fdg3$
($24\degr \la \alpha \la 36\degr $, 
$-75\fdg0 \la \delta \la -73\fdg7$)
in the $J, H$, and $K_S$ bands.
On the basis of colors and magnitudes,
about 200 Herbig Ae/Be star candidates are selected.
Considering the contaminations by miscellaneous sources
such as foreground stars and early-type dwarfs in the Magellanic Bridge,
we estimate that about 80 ($\approx 40$ \%) of the candidates are
likely to be Herbig Ae/Be stars.
We also found one concentration of the candidates
at the young star cluster NGC 796,
strongly suggesting the existence of pre-main-sequence (PMS) stars
in the Magellanic Bridge.
This is the first detection of PMS star candidates
in the Magellanic Bridge, and 
if they are genuine PMS stars,
this could be direct evidence of recent star formation.
However, the estimate of the number of Herbig Ae/Be stars depends on the fraction of
classical Be stars,
and thus a more precise determination of the Be star fraction or
observations to differentiate
between the Herbig Ae/Be stars and classical Be stars are required.

\end{abstract}



\section{INTRODUCTION}
\label{sec:Intro}

The wing of the Small Magellanic Cloud (SMC)
was detected by \citet{Shapley40},
who described it as a large elliptical extension
consisting of a cloud of blue stars.
He also implied that there is a physical connection 
between the SMC and the Large Magellanic Cloud (LMC),
since the wing extends 
from the center of the SMC in the direction of the LMC. 
A number of \ionn{H}{I} surveys
\citep[e.g.,][]{Hindman63,Muller03a}
have since revealed a continuous bridge of gas
between the SMC and LMC, now known as the Magellanic Bridge (MB).
The optical ``stellar bridge'' was found by \citet{Irwin90}
between the wing and the western halo of the LMC,
showing that
the SMC wing is the brightest section of the stellar link between them.
The wing is the richest region of 
blue stars \citep{Irwin90}, clusters/associations \citep{Bica95},
and \ionn{H}{I} gas \citep{Muller03a} in the MB,
and is thus expected to be the most active in star formation.

The MB is the closest bridge 
(60 kpc, assuming that its distance is the same as that to the SMC; 
e.g., Harries, Hilditch, \& Howarth 2003) 
among a number of known intergalactic bridges situated in groups of galaxies.
The mechanism responsible for the formation
of the MB is widely considered to be
the gravitational influence of the LMC 
\citep[e.g.,][]{Gardiner94}.
Because of tidal interactions, 
star formation is likely to be very active in the bridges,
and the details of these interactions can be investigated
through the study of young stars, such as OB and pre-main-sequence (PMS) stars.
The MB can also provide us with
insights into the evolutionary development of dwarf galaxies
as well as the role of external dynamical interactions
in stimulating star formation.
The relative importance of the star formation processes
due to external and internal disturbances 
to the SMC is one of the key questions
yet to be properly answered.

Another important aspect of observations is that
stars in the MB can provide
valuable clues for understanding
star formation processes in a low metallicity environment.
The very low metallicity of the MB was found
to be $\sim 0.5$ dex lower than that of the SMC \citep{Rolleston99},
which itself is much lower than our Galaxy.
De Wit et al. (2003) discussed properties of
PMS (Herbig Ae/Be) stars in the SMC
and suggested that such stars have luminosities generally higher
than those of the same spectral type in our Galaxy,
indicating a faster proto-stellar mass accretion rate
at lower metallicity \citep[see also;][]
{Beaulieu96,Beaulieu01,Lamers99,Panagia00,deWit02,deWit05}.
Because of their vicinity, 
the PMS stars in the MB
offer a good opportunity to study in detail 
the formation and evolution of individual young stars
in different environments.

Previous studies of the star formation processes in the MB
have mainly concentrated on stellar populations 
at optical wavelengths
\citep[e.g.,][]{Wester71,Demers98}
and CO content \citep{Muller03b,Mizuno06}.
In this paper, we present the result
of our near infrared ($J, H$, $K_S$) survey 
of the western region in the MB (the wing of the SMC).
The limiting magnitudes of the survey are $\sim 2$-$3$ mag deeper than
the Two Micron All Sky Survey (2MASS),
and this difference is critical to detecting
intermediate-mass young stellar objects (i.e., Herbig Ae/Be stars)
at the distance of the Magellanic Clouds and the MB.
The age of the Herbig Ae/Be stars we can detect is 
less than $\sim 10^{6}$ yr,
the youngest stellar population ever found in the MB.
Our detection will also
make it possible to understand star formation processes
in a low metallicity environment 
in the region far from the center of the SMC.

\section{OBSERVATIONS AND DATA REDUCTION}
\label{sec:Obs}

Our observations were conducted in 2004 August
and 2005 March, July$-$August
using the near-infrared camera SIRIUS
\citep[Simultaneous Infrared Imager for Unbiased Survey;][]{Nagas99, Nagay03} 
on the IRSF (Infrared Survey Facility) telescope.
IRSF is a 1.4 m telescope constructed and operated 
by Nagoya University and SAAO (South African Astronomical Observatory)
at Sutherland, South Africa.
The SIRIUS camera
can provide $J$ (1.25 $\mu$m), $H$ (1.63 $\mu$m),
and $K_S$ (2.14 $\mu$m) images simultaneously,
with a field of view of 7\farcm7 $\times$ 7\farcm7
and a pixel scale of 0\farcs45.

The total area covered by our survey
is $\sim 4$ deg$^2$
(324 fields)
over $1^{\mathrm{h}} 36^{\mathrm{m}} \la \mathrm{RA ~(J2000.0)} 
\la 2^{\mathrm{h}} 24^{\mathrm{m}} $ and
$-75\fdg0 \la \mathrm{Dec ~(J2000.0)} \la -73\fdg7$ 
(Fig. \ref{fig:HIHAeBeSpat}).
We observed only on photometric nights,
and the typical seeing was 1\farcs1 FWHM in the $H$ band.
A single image comprises 10 dithered 30 s exposures.

Data reduction was carried out with 
the IRAF (Image Reduction and Analysis Facility)\footnote{
IRAF is distributed by the National Optical Astronomy
Observatory, which is operated by the Association of Universities for
Research in Astronomy, Inc., under cooperative agreement with
the National Science Foundation.}
software package.
Images were prereduced following the standard procedures
of near-infrared arrays 
(dark frame subtraction, flat-fielding, and sky subtraction).
Photometry, including point-spread function (PSF) fitting, was carried out 
with the DAOPHOT package \citep{Stetson87}.
We used the DAOFIND task to identify point sources,
and the sources were then utilized 
for PSF-fitting photometry in the ALLSTAR task.
Each image was calibrated with the standard stars
9103, 9106, and 9109 \citep{Persson98}.
The averages of the 10$\sigma$ limiting magnitudes were
$J=19.0$, $H=18.0$, and $K_S=17.2$.
The averages of the zero-point uncertainties
were about 0.03, 0.03, and 0.04 mag 
in the $J,H$ and $K_S$ bands, respectively.
We also confirmed that
the magnitude difference between our data and 2MASS
is 0.02-0.04 mag in the three bands.
This data set is the preliminary version of 
``The IRSF Magellanic Clouds Point Source Catalog'';
a more detailed description of our analysis 
is given in D. Kato et al. (2007, in preparation).

\section{SELECTION OF HERBIG Ae/Be STARS}
\label{sec:Results}

In our fields, a total of 49,865, 34,964, and 22,094 stars are detected
in the $J,H$ and $K_S$ bands, respectively, 
within the 10 $\sigma$ limiting magnitude.
To find PMS stars with infrared excess,
we made a $J-H$ versus $H-K_S$ color-color (CC) diagram 
(Fig. \ref{fig:CCM}).
Only 20,261 stars detected in all three bands with photometric errors 
$\sigma_{\mathrm{phot}} \leq 0.1$
mag are plotted.
The CC diagram of the MB shows three distinct features.
First, most ($\sim 94$ \%) of the stars are distributed
around dwarf and giant loci of F to M spectral types
($0.2 \la J-H \la 0.8$, $-0.2 \la H-K_S \la 0.5$).
They are dominated by foreground sources,
and their classification into foreground and Bridge sources
is not straightforward.
Second, some sources ($\sim 3$ \%, $\sim 800$) are located at 
$0.5 \la J-H \la 1.2$ and $0.5 \la H-K_S \la 1.0$.
Most of them are fainter than 15.0 mag in the $J$ band.
There could be three populations in this area:
group I/I\hspace{-.1em}I Herbig Ae/Be stars \citep{Hillenb92},
AGB stars, and predominantly galaxies \citep[cf.][]{Nakaj05}.
Third, we can find some sources 
at $J-H \la 0.2$ and $H-K_S \la 0.4$.
They are located on and rightward of the dwarf locus,
indicating that they are normal early type (OB) dwarfs
and stars with infrared excess.

Herbig Ae/Be stars in our Galaxy
show strong excess radiation
at infrared wavelengths.
This excess is interpreted in terms of circumstellar dust emission
with a wide range in temperatures.
For our sources redder than $J-H = 0.2$,
selection of the Herbig Ae/Be stars is complicated
due to contamination by dwarfs/giants and background galaxies
(see Fig. \ref{fig:CCM}).
Since the Herbig Ae/Be stars in the LMC and SMC
have a bluer color than those in our Galaxy 
\citep[e.g.,][]{deWit03,deWit05},
we searched 
the region bluer than $J-H = 0.2$,
where stars with small infrared excess
are located,\footnote{
In this analysis, we regard
the interstellar extinction as negligible.
The largest H {\scriptsize I} column density in our region
is $\sim 1.5 \times 10^{21}$ cm$^{-2}$,
corresponding to $0.79$ mag in the $V$ band
and $0.064$ mag in the $K$ band for solar metallicity
($N_H/A_V=1.9 \times 10^{21}$ cm$^{-2}$ mag$^{-1}$
[Bohlin, Savage, \& Drake 1978],
$A_K = 0.081 \times A_V$ [van de Hulst 1946]).
Moreover, the metallicity of the MB is much lower than the solar one,
hence extinction is expected to be much smaller than
$0.064$ mag in the $K$ band.
}
for the Herbig Ae/Be stars.

We selected Herbig Ae/Be candidates
with the following criteria:
(1) $J-H \leq 0.2$,
(2) $J \geq 13.4$,
(3) more than $0.1$ mag apart from the OB locus,
and (4) located under the line $J-H = 4.2 \times (H-K_S) - 0.23$.
Stars with $J-H < -0.1$ and $H-K_S > 0.4$
were also excluded.
The first criterion is, as discussed above,
to exclude Galactic late-type (later than F5) dwarfs/giants 
and galaxies.
Early type stars in our Galaxy and supergiants in the MB
are brighter than O3 V stars in the MB,
and thus the second criterion was imposed to exclude such stars.
With the third one, the OB dwarfs were rejected 
(see circles in Fig. \ref{fig:CCM}).

The fourth criterion is imposed to find stars with infrared excess.
Fig. \ref{fig:Dist} shows a histogram of the distances $d_\mathrm{locus}$
from the dwarf locus at $0.0 \leq J-H \leq 0.2$,
for the sources in the area ``A'' shown in Fig. \ref{fig:CCM}.
The locus in the area corresponds to A0$-$F5 dwarfs,
and we assume that the locus is a linear line
with $J-H = 4.2 \times (H-K_S) -0.02$.
We also assume that sources at $d_\mathrm{locus} < 0$ 
(the left side of the locus in the area A)
are AF dwarfs, 
and those at $d_\mathrm{locus} > 0$ are AF dwarfs \textit{plus} 
Herbig Ae/Be stars with the infrared excess.
The AF dwarfs are expected to show a Gaussian distribution,
$N(d_\mathrm{locus}) = N_0 \exp(- d_\mathrm{locus}^2/2\sigma^2)$.
The Gaussian distribution of AF dwarfs
determined by fitting the histogram at $d_\mathrm{locus} < 0$
is shown in Fig. \ref{fig:Dist}.
The standard deviation of the fitting is $\sigma = 0.057$ mag
and $N_0 = 23.6$.
We assume that the sources with 
$d_\mathrm{locus} \geq 0.057 ~\mathrm{mag} = 1\sigma$ 
have infrared excess.

\begin{figure}[t]
  \begin{center}
   \epsscale{0.9}
   \caption[]{
   see 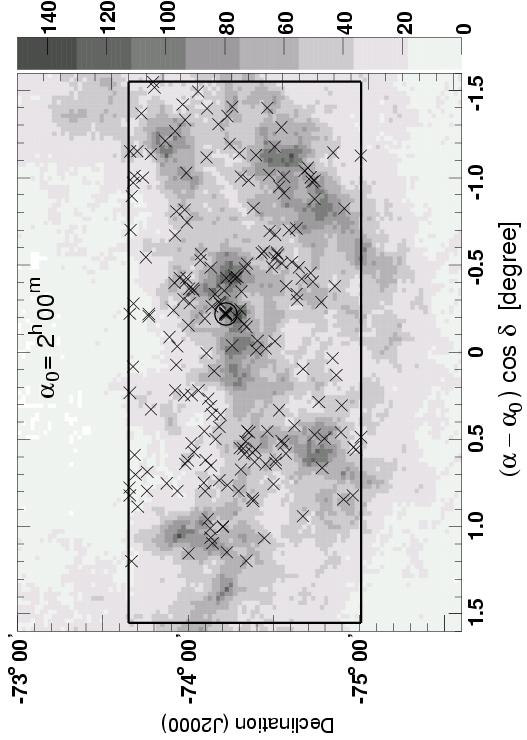 --- Herbig Ae/Be candidates (crosses)
   within the western Magellanic Bridge
   (the wing of the SMC)
   found in this survey
   overlaid on an integrated \ionn{H}{I} intensity map
   \citep{Muller03a}.
   Our observed area is shown by solid lines.
   The gray-scale is a linear transfer function,
   as shown on the intensity wedge 
   with units in K km s$^{-1}$.
   The position of the star cluster NGC 796
   ($01^{\mathrm{h}} 56^{\mathrm{m}}$, $-74\degr 13\arcmin$)
   is shown by a circle.
   }
    \label{fig:HIHAeBeSpat}
  \end{center}
\end{figure}

\begin{figure}[t]
 \begin{center}
  \caption[]{
   see 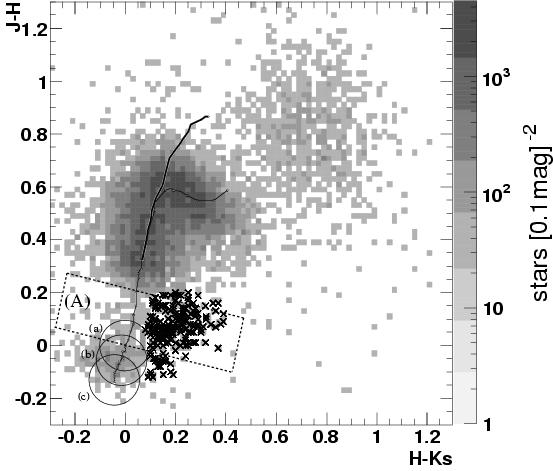 ---  $J-H$ versus $H-K_S$ color-color diagram 
  for the point sources detected in all bands
  with photometric error less than 0.1 mag (gray-scale).
  Crosses represent Herbig Ae/Be candidates.
  The thin and thick curves are 
  the loci of dwarfs and giants, respectively
  \citep{Tokunaga00}.
  Their colors are transformed to the IRSF system
  with the color equations of \citet{Nakaj05}. 
  The stars in the region labeled A
  are used to estimate the contamination by AF dwarfs.
  The three circles are
  0.1 mag radius error circles centered at (a) B9, (b) B3, and (c) O6 dwarfs,
  which are used in the third criterion to select Herbig Ae/Be candidates
  (see text, Sec. \ref{sec:Results}).
  }
  \label{fig:CCM}
 \end{center}
  \end{figure}

\begin{figure}[h]
  \begin{center}
   \plotone{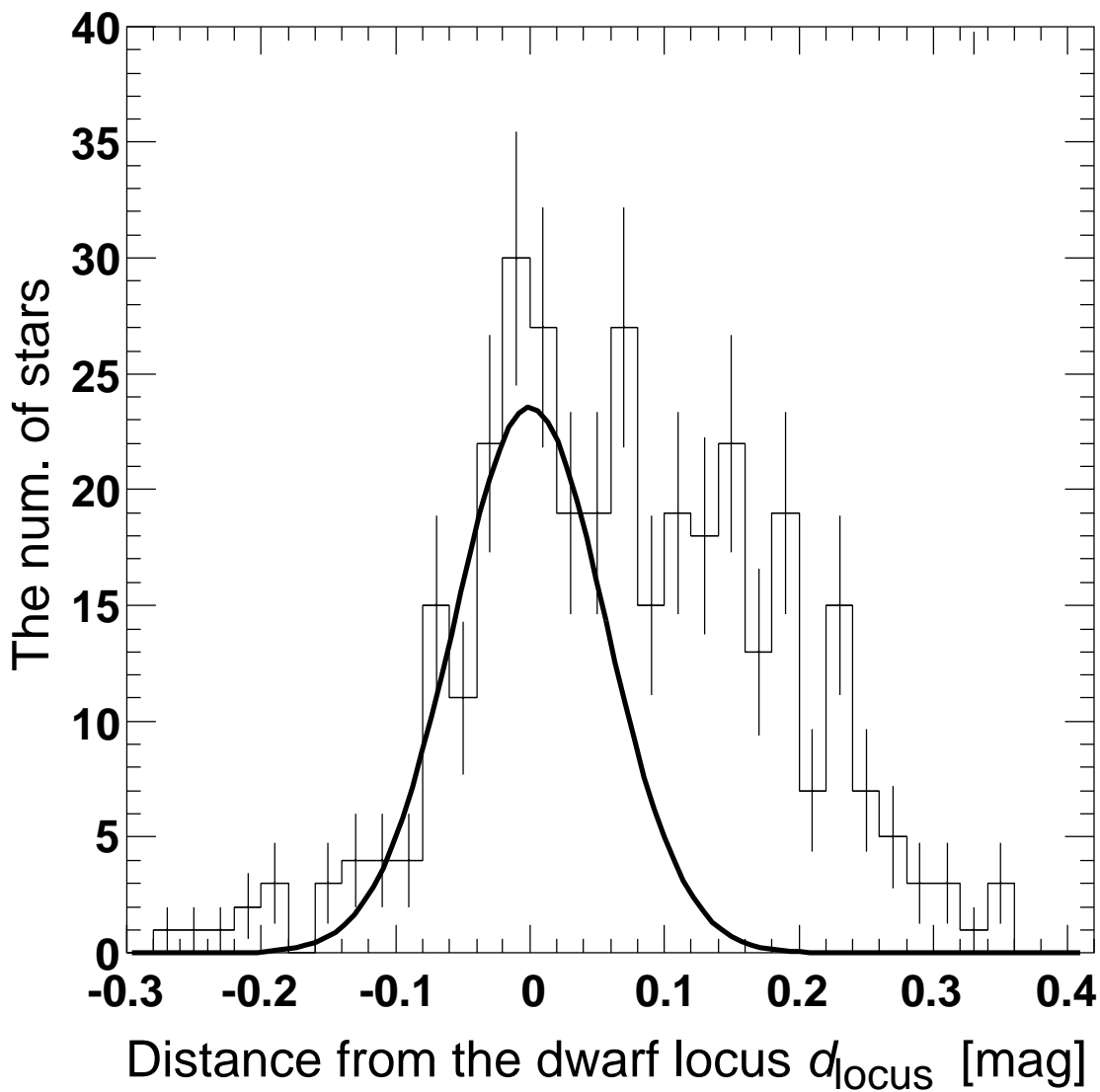}
   \hspace{0.05\linewidth}
   \caption[]{
   Distribution of distance from the dwarf A0$-$F5 locus
   for the point sources in the A region
   shown in Fig. \ref{fig:CCM}.
   The dwarf locus is assumed to be a linear line
   with $J-H = 4.2 \times (H-K_S) -0.02$.
   The thick curve denotes a Gaussian distribution
   with a mean of 0 and sigma of 0.057,
   which is derived by fitting the histogram
   at $d_{\mathrm{locus}} \leq 0$.
    }
    \label{fig:Dist}
  \end{center}
\end{figure}

As a result, we found 203 Herbig Ae/Be star candidates (Table \ref{tb}).
Their spatial distribution is shown 
in Fig. \ref{fig:HIHAeBeSpat}.
A concentration of the candidates 
associated with the star cluster NGC 796 is found
at $01^{\mathrm{h}} 56^{\mathrm{m}}$ and $-74\degr 13\arcmin$
(shown by an open circle in Fig. \ref{fig:HIHAeBeSpat}),
but we cannot find any other concentrations.
The candidates are also plotted on the CC diagram (Fig. \ref{fig:CCM}, crosses).

\section{DISCUSSION}
\label{sec:Discus}

\subsection{Contamination by Miscellaneous Sources}

There could be three populations in the region
where our Herbig Ae/Be candidates are distributed
(shown by crosses in Fig. \ref{fig:CCM}):
Herbig Ae/Be stars, classical Be (CBe) stars,
and dwarfs of O to F5 spectral types.
To confirm whether 
Herbig Ae/Be stars exist in the MB or not,
the number of the latter two populations
contaminating the region should be estimated.

The contamination by AF (A0 to F5) dwarfs can be estimated
by using the histogram shown in Fig. \ref{fig:Dist}.
As mentioned above, 
the distribution of distance from the AF dwarf locus
can be given by a Gaussian function,
and thus the number of AF dwarfs distributed at 
$d_\mathrm{locus} \geq 0.057$ is calculated to be $\approx 25$.

\begin{figure}[h]
  \begin{center}
   \plotone{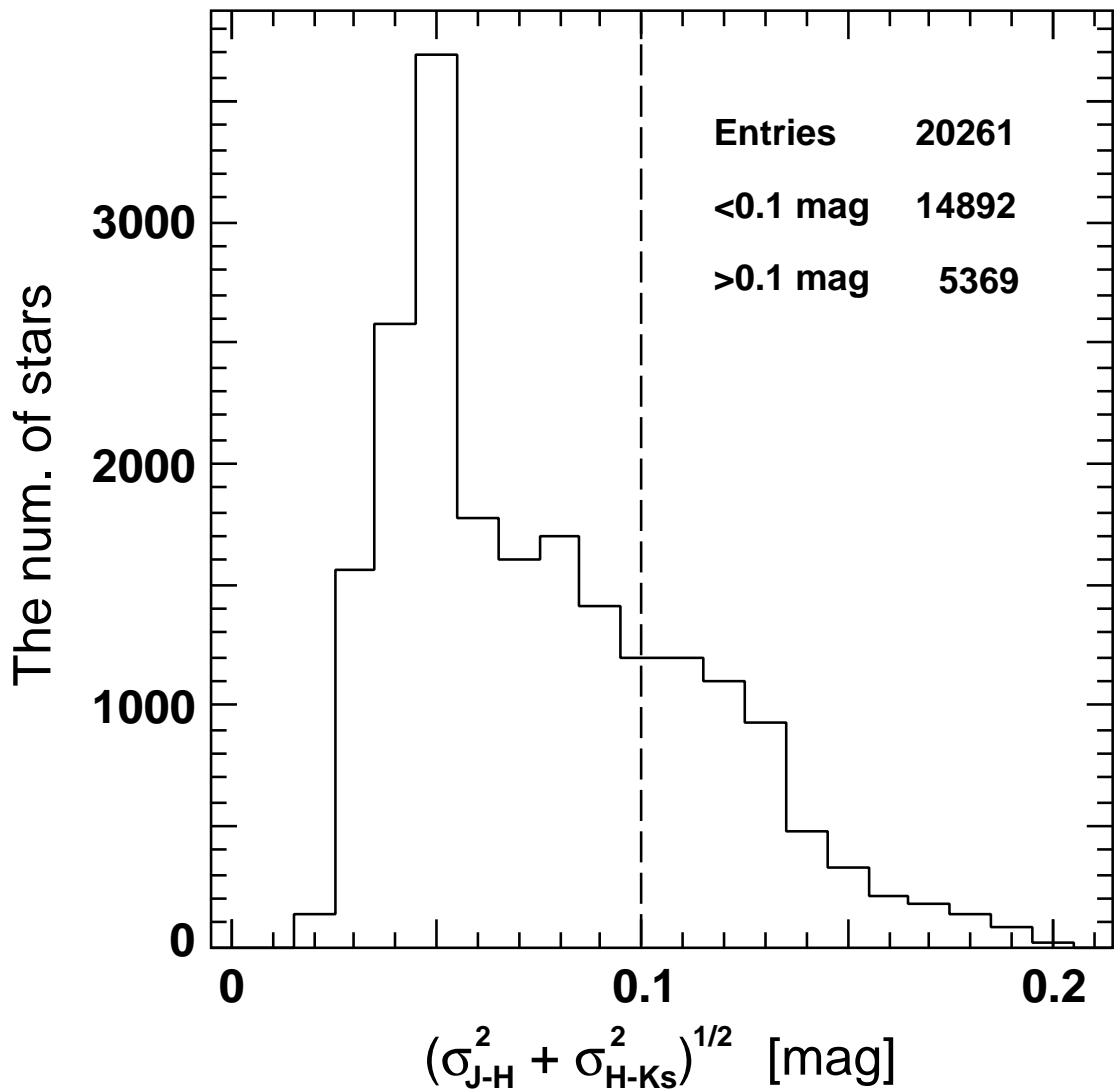}
   \hspace{0.05\linewidth}
   \caption[]{
	Histogram of $\sigma_{\mathrm{tot}} = \sqrt{\sigma^2_{J-H} + \sigma^2_{H-K_S}}$ 
	for all sources detected in the three bands 
	with $\sigma_{J,H,K_S} \leq 0.1$  mag.
	The numbers of sources with $\sigma_{\mathrm{tot}} < 0.1$
	and $\sigma_{\mathrm{tot}} > 0.1$ mag are
	14,892 (74 \%) and 5,369 (26 \%), respectively.
    }
    \label{fig:Error}
  \end{center}
\end{figure}

Since OB dwarfs are distributed around the locus
with an observational uncertainty,
we can estimate how many OB dwarfs are contaminating
the region of stars with the infrared excess.
Among the sources detected in the three bands with 
$\sigma_{\mathrm{phot}} \leq 0.1$ mag,
about 26 \% have an uncertainty 
$\sqrt{\sigma^2_{J-H} + \sigma^2_{H-K_S}} > 0.1$ mag (Fig. \ref{fig:Error}).
This means that the number of OB dwarfs distributed outside of 
the 0.1 mag radius error circles (see, Fig. \ref{fig:CCM})
is less than 26\% of the number of those within the error circles (333 stars).
Half (right side) of the outside stars are
located in the region of stars with infrared excess,
and thus contamination by OB dwarfs is less than 13\%.

CBe stars are B-type stars
that exhibit line emission over the photospheric spectrum.
The observational characteristics
of CBe stars and Herbig Ae/Be stars are very similar,
making it a very difficult task to differentiate
between the two types at near-infrared wavelengths.

In the SMC and LMC fields, 
the mean number ratio of CBe stars to early B-type stars\footnote{
Six clusters used by \citet{Keller99} are not so young (10$-$30 Myr)
that Herbig Ae/Be stars are not included in their samples of Be stars.}
is estimated to be $\sim 20 \%$ 
\citep{Keller99}.
From the 2MASS $JHK_S$ magnitudes of Galactic CBe stars\footnote{
Following \citet{Zhang05},
we regard all the stars in their list as classical Be stars,
although there are a few Herbig Ae/Be stars (see their Sec. 2). }
\citep{Zhang05}, we derived
the fraction of CBe stars in the error circles discussed above
(see also Fig. \ref{fig:CCM})
and those selected as Herbig Ae/Be stars with the criteria
from Sec. \ref{sec:Results} to be 47 \% and 48 \%, respectively,
and thus we can estimate the number of contamination by CBe and OB stars as follows.
The number of stars in the error circles, 333,
is the sum of 74 \% of OB dwarfs and 47 \% of CBe stars
($0.74 N_{\mathrm{OB}} + 0.47 N_{\mathrm{CBe}} = 333$, 
where $N_{\mathrm{OB}}$ and $N_{\mathrm{CBe}}$
are the number of OB and CBe stars, respectively).
The result of \citet{Keller99} gives $4 N_{\mathrm{CBe}} = N_{\mathrm{OB}}$,
and thus we obtain $N_{\mathrm{CBe}} \approx 97$ and $N_{\mathrm{OB}} \approx 388$.
Therefore, the number of contaminating sources is
$0.48 N_{\mathrm{CBe}} + 0.13 N_{\mathrm{OB}} + 25 (\mathrm{AF stars}) \approx 47 + 50 + 25 = 122$,
suggesting that about 60 \% 
of the 203 Herbig Ae/Be candidates could be
contaminated by the AF and OB dwarfs, and the CBe stars.
Resultantly, we conclude that
81 of them are Herbig Ae/Be stars
after eliminating the possible contaminating sources.

However, the frequency of the CBe stars
among B-type stars is still controversial.
By using bright field stars in our Galaxy,
\citet{Zorec97} reported that
as many as one-third of early B (B0 to B4) stars show the Be phenomenon.
If we took their frequency,
a contamination of 151 stars would be found.
On the other hand, 
in the sample of Galactic open clusters,
\citet{McSwain05} showed that
Be stars comprise only 7.4 \% and 4.7 \% of
the evolved and early-type B stars, respectively.
This could lead to the estimate that
about half of the sources with infrared excess are Herbig Ae/Be stars.
Hence, a more precise determination of the CBe frequency or 
observations to differentiate
between the Herbig Ae/Be stars and CBe stars are required 
to decide how many stars have been formed in the MB.

\subsection{NGC 796}
\label{subsec:NGC796}

\begin{figure}[t]
  \begin{center}
   \epsscale{0.7}
   \plotone{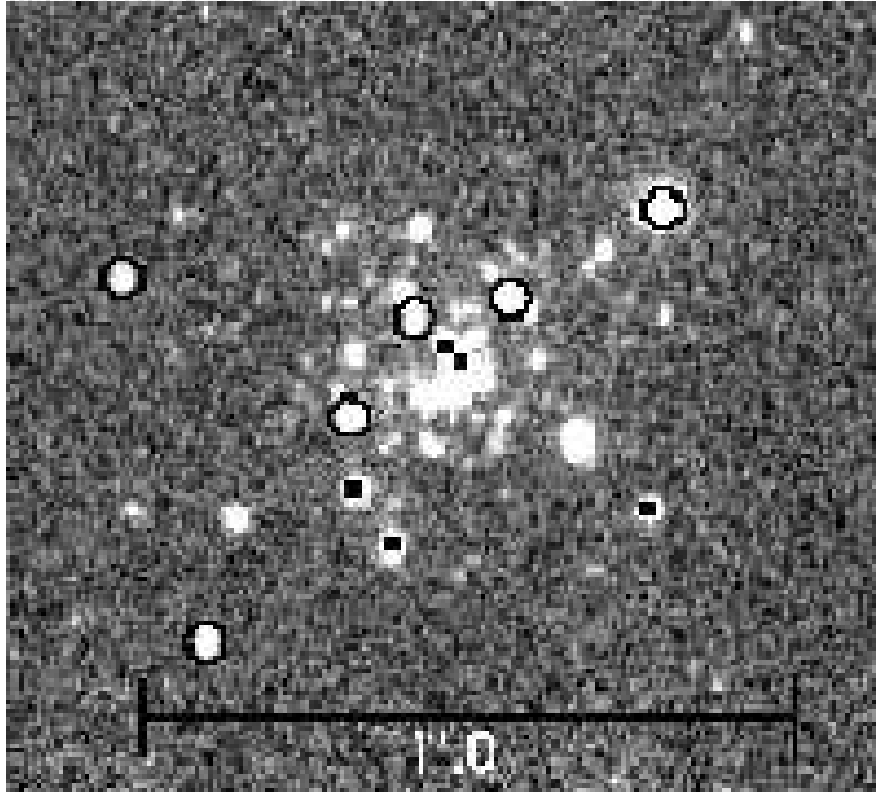}
   \caption[]{
     $J$ band image of NGC 796.
     A scale of $1\arcmin$ is shown in the image.
   North is up, and east is to the left.
   Circles and squares represent
   Herbig Ae/Be star candidates
   and stars near the OB locus, respectively.
   }
   \label{fig:NGC796}
  \end{center}
\end{figure}

NGC 796 (Fig. \ref{fig:NGC796}) is one of the youngest clusters 
in the SMC$-$Bridge region, 
and its age is estimated to be less than $\sim 10$ Myr
from the comparison of integrated spectra 
with those of template clusters \citep{Ahumada02}.
This is the only association with an \textit{IRAS} counterpart
in the inter-cloud region \citep{Grondin93}.
We have found six Herbig Ae/Be candidates
in the $\sim 30\arcsec$ radius centered at NGC 796 
(open circles in Fig. \ref{fig:NGC796}).
Five OB star candidates 
(the stars near the dwarf locus of spectral types O-B
on the CC diagram)
are also distributed in the same radius
(filled squares in Fig. \ref{fig:NGC796}).

As mentioned above,
the CBe stars could be located at
the infrared-excess area on the CC diagram;
however, clusters younger than 10 Myr are 
almost completely lacking CBe stars
\citep{Fabregat00,McSwain05}.
Considering the age of NGC 796,
most of the stars with infrared excess are likely
Herbig Ae/Be stars.
This fact further suggests that
our criteria for selecting the Herbig Ae/Be stars
are appropriate.
The \textit{IRAS} detection sensitivity for protostars 
at the distance of the MB 
(single protostars with masses $\ga 10$$-$$20 \mathrm{M}_{\sun}$
or an association containing a few less massive stars,
Grondin \& Demers 1993)
also indicates that NGC 796 is a cluster of Herbig Ae/Be stars.
We found no other strong concentrations of Herbig Ae/Be candidates
in our survey region.

\subsection{Herbig Ae/Be Stars in the Magellanic Bridge}
\label{subsec:HAeBeMB}

Higher-mass PMS stars are not expected to be visible 
before they reach the zero-age main sequence
due to obscuration by circumstellar dust envelopes/disks,
but the visibility may depend on
environmental conditions such as metallicity.
For example, \citet{Beaulieu96}
have found Herbig Ae/Be candidates in the LMC
with luminosities systematically higher 
than observed in our Galaxy.
The candidates are located well above the birthline of \citet{Palla93}
for an accretion rate of $10^{-5} \mathrm{M_\sun~yr^{-1}}$
in the HR diagram.
\citet{Beaulieu96} then concluded that 
their location in the HR diagram indicates
a lower circumstellar extinction due to low metallicity, or
a prevailing accretion rate of about $10^{-4} \mathrm{M_\sun~yr^{-1}}$,
10 times higher than for Galactic stars 
of the same luminosity.
\citet{Panagia00} suggested
a higher accretion rate for LMC low-mass PMS stars as well,
and recently 
a statistically significant excess of the accretion rate
($\sim 30$ times more than that for Galactic stars at a comparable age)
was found
for low-mass PMS stars in the LMC \citep{Romaniello04}.
However, the location of the birthline could also be controlled
by the deuterium mass fraction and stellar opacity,
which are modified by changing the metallicity 
\citep{Bernasconi96alone}.

\begin{figure}[t]
 \begin{center}
  \epsscale{1.0}
 \end{center}
 \caption[]{
see 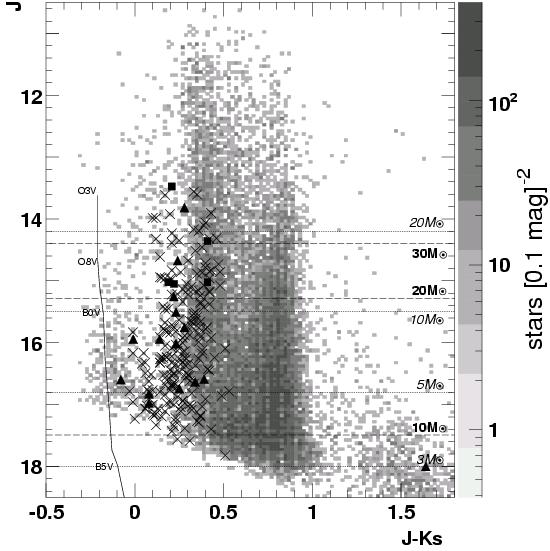 ---  $J$ versus $J-K_S$ color-magnitude diagram 
 for point sources detected in all bands,
 with photometric error less than 0.1 mag (gray-scale).
 Crosses represent Herbig Ae/Be candidates
 found in our survey,
 and triangles and squares
 represent the Herbig Ae/Be candidates
 found in the LMC \citep{deWit05}
 and in the SMC \citep{deWit03}, respectively.
 The solid curve is
 the locus of dwarfs \citep{Tokunaga00}.
 The dotted and dashed lines
 denote the masses of 
 the group I/I\hspace{-.1em}I and
 the group I\hspace{-.1em}I\hspace{-.1em}I Herbig Ae/Be stars, respectively,
 estimated by the empirical relationship
 between $J$ magnitude and stellar mass
 \citep{Hillenb92,Nakaj05}
 at the distance of the MB (60 kpc).
 Their colors are transformed to the IRSF system
 with the color equations of \citet{Nakaj05}. 
 }
 \label{fig:CMD}
\end{figure}

Our result suggests that
the environmental effect of low metallicity
can also be seen for the PMS stars in the MB.
Fig. \ref{fig:CMD} shows the location of 
the Herbig Ae/Be candidates in the color magnitude (CM) diagram (crosses).
Assuming that our candidates are
the group I\hspace{-.1em}I\hspace{-.1em}I Herbig Ae/Be stars
defined by \citet{Hillenb92},
the $J$ band magnitudes show that 
the candidates have masses in the range of 
$\sim 10$ to more than 30 $\mathrm{M_\sun}$
(see horizontal dashed lines in Fig. \ref{fig:CMD}).
If they are members of the group I/I\hspace{-.1em}I,
the mass range is $\sim 3$ to more than $20 \mathrm{M_\sun}$
(horizontal dotted lines in Fig. \ref{fig:CMD}).
The Herbig Ae/Be candidates
found in the LMC (filled triangles, de Wit et al. 2005)
and in the SMC (filled squares,  de Wit et al. 2003)
are also plotted in Fig. \ref{fig:CMD}.
They are up to 10 times as luminous as 
the Galactic Herbig Ae/Be stars in the same temperature range
\citep[see, e.g.,][Sec. 9.2 and Fig. 18]{deWit03},
and both our candidates and theirs are located in a region
where a few Galactic Herbig Ae/Be stars are found.\footnote{
We have adopted a distance to the MB of 60 kpc.
If we adopted 50 kpc, which is the distance to the LMC, 
the stars we detected would be fainter by only 0.4 mag,
but the Herbig Ae/Be candidates would still be
brighter than the Galactic average.}
The location of our candidates in the CM diagram
thus implies that
the formation process of the Herbig Ae/Be stars in the MB, 
as well as the LMC and SMC,
is different from that in our Galaxy.

The MB is believed to have been generated by
the recent collision of the Magellanic Clouds
about 200 Myr ago \citep{Gardiner94}.
The existence of the Herbig Ae/Be stars, associations 
and other young massive stars
($<20$ Myr, e.g., Demers \& Battinelli 1998, Rolleston et al. 1999) 
indicates that star formation is still occurring in the MB.

In the MB, we found $\sim 80$ Herbig Ae/Be stars,
whose smallest masses are about 
$3 \mathrm{M_\sun}$ (group I/I\hspace{-.1em}I) and
$10 \mathrm{M_\sun}$ (group I\hspace{-.1em}I\hspace{-.1em}I)
(see, Fig. \ref{fig:CMD}).
According to the calculation by \citet{Bernasconi96},
contraction times deduced for 
$3$ and $9 \mathrm{M_\sun}$ stars at $Z = 0.001$
are $\approx 1.7$ and $0.2$ Myr, respectively.
We can thus estimate, to the first order, 
the {\it lower limit} of 
the star formation rate in the MB in the last megayear
as $\sim 1.4 \times 10^{-4}~ \mathrm{M_\sun~yr^{-1}}$
for group I/I\hspace{-.1em}I.
This rate is comparable to 
$\sim 4 \times 10^{-4}~ \mathrm{M_\sun~yr^{-1}}$,
derived from the observation of $^{12}$CO(1-0) emission \citep{Mizuno06}.

How long the MB will survive as an independent system
is as yet uncertain.
The gas in the MB shows low velocities
in the LMC-standard-of-rest frame,
indicating that the gas will certainly be absorbed
into the Magellanic Clouds \citep{Bruns05}.
However,
more than $10^{5}~ \mathrm{M_\sun}$
stars could be formed
in a few gigayears \citep{Yoshizw03} if it keeps its current star formation rate.
This implies that the MB has the potential to evolve into a dwarf galaxy.

\section{CONCLUSION}
\label{sec:Conc}

We have conducted a near infrared ($J,H,K_S$) deep survey 
of the western region of the Magellanic Bridge 
(the wing of the SMC).
Using color-color and color-magnitude diagrams,
we have found $\sim 200$ Herbig Ae/Be star candidates.
Considering contamination by other sources,
we infer that about 40 \% of them are Herbig Ae/Be stars
in the Magellanic Bridge.
We have also found six Herbig Ae/Be star and five OB star candidates 
in NGC 796,
showing the existence of a very young cluster in the region.
The location of the Herbig Ae/Be candidates 
in the color-magnitude diagram
indicates that they are more luminous than the Galactic ones.

\acknowledgements

We are grateful to Eric Muller for kindly providing the \ionn{H}{I} data
of the Magellanic Bridge.
We would like to thank Haruyuki Okuda, Norikazu Mizuno, and Akiko Kawamura
for their helpful comments.
We also thank the staff at the South African Astronomical Observatory (SAAO)
for their support during our observations.
The IRSF/SIRIUS project was initiated and supported by Nagoya
University, the National Astronomical Observatory of Japan
in collaboration with the SAAO
under Grants-in-Aid for Scientific Research
No.10147207, No.10147214, No.13573001, and No.15340061
of the Ministry of Education,
Culture, Sports, Science and Technology (MEXT) of Japan.
This work was also supported in part 
by the Grants-in-Aid for the 21st Century 
COE ``The Origin of the Universe and Matter: 
Physical Elucidation of the Cosmic History'' from the MEXT of Japan.


\clearpage
\begin{deluxetable}{rcccccccc}
\tablewidth{0pt}
\tablecaption{Catalog of Herbig Ae/Be Star Candidates in the Magellanic Bridge \label{tb}}
\tablehead{
\colhead{ID}           
& \colhead{R.A. (J2000.0)} & \colhead{Dec. (J2000.0)}
& \colhead{$J$}  & \colhead{$\sigma_{J}$}
& \colhead{$H$}  & \colhead{$\sigma_{H}$}  
& \colhead{$K_S$}  & \colhead{$\sigma_{K_S}$} }
\startdata
1 & 01 37 48.37 & $-$73 47 42.2 & 16.18 & 0.02 & 16.04 & 0.03 & 15.80 & 0.04\\
2 &  01 38 13.75 & $-$73 48 08.9 &  16.35 &  0.02 &  16.31 &  0.03 &  16.12 &  0.04\\
3 &  01 38 15.48 & $-$74 03 27.7 &  16.62 &  0.03 &  16.47 &  0.05 &  16.31 &  0.06\\
4 &  01 39 05.58 & $-$74 27 47.0 &  16.79 &  0.03 &  16.63 &  0.06 &  16.26 &  0.08\\
5 &  01 39 15.82 & $-$74 15 32.3 &  15.91 &  0.02 &  15.74 &  0.09 &  15.54 &  0.02\\
6 &  01 39 26.63 & $-$73 58 02.9 &  16.22 &  0.02 &  16.15 &  0.02 &  16.01 &  0.04\\
7 &  01 39 34.99 & $-$74 06 39.9 &  13.56 &  0.03 &  13.47 &  0.02 &  13.23 &  0.06\\
8 &  01 40 05.62 & $-$74 13 40.8 &  14.19 &  0.02 &  14.20 &  0.03 &  13.83 &  0.03\\
9 &  01 40 29.37 & $-$73 43 28.0 &  15.51 &  0.02 &  15.36 &  0.02 &  15.17 &  0.03\\
10 &  01 40 39.68 & $-$74 32 43.5 &  17.03 &  0.04 &  16.89 &  0.06 &  16.65 &  0.09\\
11 &  01 40 42.83 & $-$73 58 57.0 &  17.27 &  0.03 &  17.15 &  0.05 &  16.92 &  0.10\\
12 &  01 40 50.80 & $-$74 10 30.9 &  15.16 &  0.02 &  15.03 &  0.02 &  14.80 &  0.02\\
13 &  01 41 42.49 & $-$73 55 28.8 &  15.53 &  0.01 &  15.42 &  0.02 &  15.20 &  0.03\\
14 &  01 42 21.87 & $-$74 30 14.0 &  16.69 &  0.02 &  16.63 &  0.05 &  16.47 &  0.09\\
15 &  01 42 26.35 & $-$74 14 32.2 &  13.60 &  0.01 &  13.52 &  0.07 &  13.25 &  0.02\\
16 &  01 42 29.18 & $-$74 50 46.8 &  15.05 &  0.01 &  14.92 &  0.02 &  14.76 &  0.02\\
17 &  01 42 33.64 & $-$73 51 56.9 &  15.46 &  0.02 &  15.41 &  0.02 &  15.29 &  0.04\\
18 &  01 42 34.35 & $-$75 00 33.8 &  17.18 &  0.02 &  16.99 &  0.04 &  16.81 &  0.08\\
19 &  01 43 09.84 & $-$74 18 00.1 &  16.09 &  0.02 &  15.98 &  0.02 &  15.88 &  0.04\\
20 &  01 43 11.35 & $-$74 23 50.2 &  16.14 &  0.02 &  16.05 &  0.02 &  15.86 &  0.04\\
21 &  01 43 35.61 & $-$73 42 12.7 &  16.52 &  0.03 &  16.45 &  0.02 &  16.36 &  0.05\\
22 &  01 43 39.22 & $-$73 39 31.6 &  17.56 &  0.04 &  17.49 &  0.06 &  17.32 &  0.09\\
23 &  01 43 43.00 & $-$74 06 27.5 &  17.18 &  0.03 &  17.25 &  0.05 &  17.07 &  0.07\\
24 &  01 43 44.74 & $-$73 46 56.4 &  14.98 &  0.02 &  14.89 &  0.02 &  14.77 &  0.03\\
25 &  01 43 59.29 & $-$74 41 59.5 &  13.92 &  0.01 &  13.88 &  0.01 &  13.72 &  0.02\\
26 &  01 44 15.29 & $-$74 40 54.1 &  16.24 &  0.02 &  16.17 &  0.02 &  16.06 &  0.04\\
27 &  01 44 47.42 & $-$74 43 00.4 &  16.15 &  0.01 &  16.01 &  0.02 &  15.88 &  0.04\\
28 &  01 44 48.28 & $-$74 28 22.3 &  14.95 &  0.01 &  14.93 &  0.09 &  14.80 &  0.02\\
29 &  01 44 53.65 & $-$74 33 36.2 &  13.90 &  0.04 &  13.83 &  0.05 &  13.47 &  0.02\\
30 &  01 45 04.09 & $-$74 44 18.2 &  15.75 &  0.01 &  15.64 &  0.02 &  15.53 &  0.03\\
31 &  01 45 05.01 & $-$73 59 20.1 &  15.07 &  0.01 &  14.97 &  0.01 &  14.68 &  0.02\\
32 &  01 45 07.23 & $-$74 18 32.5 &  16.80 &  0.02 &  16.74 &  0.03 &  16.66 &  0.06\\
33 &  01 45 10.67 & $-$74 43 26.7 &  14.61 &  0.01 &  14.56 &  0.01 &  14.47 &  0.02\\
34 &  01 45 25.11 & $-$74 21 06.3 &  13.61 &  0.01 &  13.53 &  0.01 &  13.44 &  0.02\\
35 &  01 45 35.89 & $-$73 41 03.2 &  15.92 &  0.02 &  15.76 &  0.02 &  15.49 &  0.03\\
36 &  01 45 43.78 & $-$73 44 03.1 &  15.99 &  0.02 &  15.86 &  0.02 &  15.63 &  0.03\\
37 &  01 45 47.08 & $-$74 31 53.2 &  16.34 &  0.03 &  16.45 &  0.03 &  16.30 &  0.06\\
38 &  01 46 11.29 & $-$73 40 43.7 &  17.82 &  0.03 &  17.70 &  0.06 &  17.31 &  0.10\\
39 &  01 46 18.17 & $-$74 33 37.4 &  16.53 &  0.02 &  16.38 &  0.04 &  16.22 &  0.06\\
40 &  01 46 39.84 & $-$74 44 19.6 &  14.06 &  0.07 &  13.94 &  0.06 &  13.66 &  0.09\\
41 &  01 47 15.78 & $-$73 40 09.3 &  16.25 &  0.02 &  16.20 &  0.02 &  16.09 &  0.04\\
42 &  01 47 21.50 & $-$74 54 43.2 &  15.17 &  0.01 &  15.00 &  0.08 &  14.75 &  0.02\\
43 &  01 47 44.04 & $-$74 22 52.0 &  17.11 &  0.03 &  17.09 &  0.05 &  16.94 &  0.09\\
44 &  01 48 10.27 & $-$73 58 45.7 &  15.66 &  0.02 &  15.54 &  0.02 &  15.34 &  0.04\\
45 &  01 48 19.98 & $-$73 55 50.0 &  15.92 &  0.01 &  15.85 &  0.02 &  15.65 &  0.04\\
46 &  01 49 11.89 & $-$73 59 31.3 &  14.86 &  0.02 &  14.79 &  0.02 &  14.60 &  0.02\\
47 &  01 49 18.59 & $-$74 37 53.8 &  15.79 &  0.02 &  15.75 &  0.02 &  15.62 &  0.04\\
48 &  01 49 29.37 & $-$74 35 22.6 &  15.78 &  0.02 &  15.75 &  0.02 &  15.62 &  0.03\\
49 &  01 49 36.72 & $-$74 28 38.1 &  14.39 &  0.02 &  14.34 &  0.06 &  14.18 &  0.06\\
50 &  01 49 54.88 & $-$74 30 34.1 &  16.68 &  0.03 &  16.62 &  0.05 &  16.51 &  0.09\\
51 &  01 50 01.41 & $-$73 39 44.2 &  14.86 &  0.07 &  14.67 &  0.02 &  14.38 &  0.02\\
52 &  01 50 20.13 & $-$73 55 20.5 &  16.23 &  0.02 &  16.22 &  0.02 &  16.05 &  0.04\\
53 &  01 51 23.50 & $-$74 26 37.1 &  15.76 &  0.02 &  15.69 &  0.02 &  15.40 &  0.03\\
54 &  01 51 27.42 & $-$74 31 06.4 &  16.68 &  0.02 &  16.63 &  0.03 &  16.47 &  0.07\\
55 &  01 51 32.68 & $-$74 25 50.1 &  17.15 &  0.02 &  17.19 &  0.07 &  17.03 &  0.08\\
56 &  01 51 38.41 & $-$74 31 26.3 &  15.34 &  0.06 &  15.19 &  0.04 &  15.07 &  0.08\\
57 &  01 51 48.02 & $-$74 04 26.7 &  15.88 &  0.02 &  15.83 &  0.03 &  15.57 &  0.04\\
58 &  01 51 56.94 & $-$74 30 26.9 &  17.00 &  0.02 &  17.00 &  0.09 &  16.90 &  0.09\\
59 &  01 52 12.68 & $-$74 36 21.9 &  16.67 &  0.03 &  16.53 &  0.03 &  16.34 &  0.06\\
60 &  01 52 13.97 & $-$73 45 03.6 &  16.50 &  0.02 &  16.37 &  0.02 &  16.21 &  0.05\\
61 &  01 52 19.41 & $-$74 04 17.4 &  16.88 &  0.03 &  16.83 &  0.05 &  16.64 &  0.09\\
62 &  01 52 19.69 & $-$74 32 48.1 &  17.23 &  0.03 &  17.33 &  0.05 &  17.23 &  0.10\\
63 &  01 52 28.27 & $-$74 20 40.8 &  16.51 &  0.02 &  16.57 &  0.03 &  16.42 &  0.05\\
64 &  01 52 38.38 & $-$74 38 52.2 &  14.83 &  0.02 &  14.76 &  0.01 &  14.62 &  0.03\\
65 &  01 52 41.21 & $-$74 29 22.8 &  16.96 &  0.02 &  17.00 &  0.04 &  16.76 &  0.09\\
66 &  01 52 53.74 & $-$74 19 48.3 &  17.07 &  0.02 &  17.12 &  0.04 &  17.01 &  0.09\\
67 &  01 52 55.76 & $-$74 07 56.3 &  17.07 &  0.02 &  17.15 &  0.04 &  17.01 &  0.10\\
68 &  01 53 07.43 & $-$74 42 54.8 &  16.98 &  0.02 &  16.93 &  0.04 &  16.80 &  0.09\\
69 &  01 53 32.17 & $-$74 16 15.9 &  14.32 &  0.01 &  14.31 &  0.01 &  14.20 &  0.02\\
70 &  01 53 37.83 & $-$74 17 13.4 &  16.26 &  0.02 &  16.19 &  0.03 &  16.01 &  0.06\\
71 &  01 53 37.93 & $-$74 39 51.2 &  16.32 &  0.02 &  16.28 &  0.02 &  16.18 &  0.04\\
72 &  01 53 44.83 & $-$74 13 18.2 &  15.66 &  0.02 &  15.58 &  0.02 &  15.29 &  0.03\\
73 &  01 53 46.73 & $-$74 16 22.7 &  16.76 &  0.02 &  16.81 &  0.06 &  16.69 &  0.10\\
74 &  01 53 49.07 & $-$73 59 16.3 &  16.48 &  0.02 &  16.40 &  0.02 &  16.23 &  0.05\\
75 &  01 53 49.85 & $-$73 56 56.1 &  14.46 &  0.02 &  14.34 &  0.02 &  14.08 &  0.02\\
76 &  01 53 50.50 & $-$74 43 40.0 &  16.61 &  0.02 &  16.55 &  0.03 &  16.43 &  0.06\\
77 &  01 54 10.53 & $-$73 58 58.5 &  17.22 &  0.05 &  17.16 &  0.03 &  16.99 &  0.08\\
78 &  01 54 13.38 & $-$74 51 23.6 &  16.84 &  0.03 &  16.76 &  0.04 &  16.67 &  0.08\\
79 &  01 54 14.81 & $-$73 55 12.5 &  17.33 &  0.02 &  17.36 &  0.04 &  17.21 &  0.10\\
80 &  01 54 18.50 & $-$74 34 23.9 &  16.51 &  0.02 &  16.41 &  0.03 &  16.29 &  0.05\\
81 &  01 54 43.21 & $-$74 01 36.5 &  17.07 &  0.02 &  17.07 &  0.04 &  16.94 &  0.10\\
82 &  01 54 43.69 & $-$74 00 13.2 &  17.16 &  0.03 &  17.21 &  0.04 &  17.06 &  0.09\\
83 &  01 54 49.63 & $-$74 12 38.5 &  16.24 &  0.02 &  16.04 &  0.09 &  15.79 &  0.04\\
84 &  01 54 51.65 & $-$74 01 50.2 &  15.83 &  0.01 &  15.80 &  0.02 &  15.65 &  0.04\\
85 &  01 54 51.73 & $-$74 18 40.8 &  16.00 &  0.02 &  15.90 &  0.02 &  15.73 &  0.04\\
86 &  01 54 57.60 & $-$74 37 57.9 &  16.71 &  0.02 &  16.70 &  0.04 &  16.50 &  0.06\\
87 &  01 55 06.93 & $-$74 09 24.5 &  16.49 &  0.02 &  16.43 &  0.02 &  16.27 &  0.05\\
88 &  01 55 30.19 & $-$74 38 14.1 &  16.27 &  0.01 &  16.24 &  0.03 &  16.13 &  0.05\\
89 &  01 55 35.87 & $-$73 58 22.6 &  15.09 &  0.02 &  14.99 &  0.01 &  14.77 &  0.02\\
90 &  01 55 37.87 & $-$74 46 43.2 &  15.18 &  0.02 &  15.07 &  0.02 &  14.84 &  0.03\\
91 &  01 55 47.27 & $-$74 10 55.1 &  16.86 &  0.02 &  16.73 &  0.05 &  16.37 &  0.05\\
92 &  01 55 53.77 & $-$74 11 09.1 &  16.03 &  0.03 &  15.94 &  0.02 &  15.80 &  0.04\\
93 &  01 55 55.01 & $-$74 00 30.6 &  14.71 &  0.02 &  14.51 &  0.02 &  14.31 &  0.02\\
94 &  01 56 02.23 & $-$73 40 45.2 &  17.29 &  0.03 &  17.22 &  0.05 &  16.93 &  0.08\\
95 &  01 56 08.95 & $-$74 08 19.8 &  15.37 &  0.02 &  15.23 &  0.02 &  15.00 &  0.03\\
96 &  01 56 22.95 & $-$74 18 18.1 &  16.34 &  0.03 &  16.25 &  0.04 &  16.04 &  0.07\\
97 &  01 56 31.73 & $-$73 54 50.9 &  16.94 &  0.03 &  16.91 &  0.04 &  16.67 &  0.07\\
98 &  01 56 39.51 & $-$74 12 55.4 &  13.98 &  0.02 &  14.01 &  0.02 &  13.86 &  0.02\\
99 &  01 56 42.99 & $-$74 13 03.6 &  16.13 &  0.02 &  16.12 &  0.02 &  15.89 &  0.05\\
100 &  01 56 45.11 & $-$74 13 05.4 &  16.77 &  0.03 &  16.83 &  0.04 &  16.67 &  0.06\\
101 &  01 56 46.59 & $-$74 13 14.6 &  16.56 &  0.02 &  16.55 &  0.04 &  16.27 &  0.07\\
102 &  01 56 49.79 & $-$74 13 35.3 &  16.45 &  0.02 &  16.37 &  0.03 &  16.03 &  0.05\\
103 &  01 56 51.73 & $-$74 13 01.8 &  16.79 &  0.02 &  16.91 &  0.04 &  16.82 &  0.08\\
104 &  01 56 51.77 & $-$73 46 12.8 &  15.84 &  0.02 &  15.80 &  0.02 &  15.64 &  0.03\\
105 &  01 56 52.50 & $-$73 39 25.7 &  15.90 &  0.01 &  15.84 &  0.02 &  15.76 &  0.04\\
106 &  01 56 58.74 & $-$74 16 51.3 &  15.59 &  0.02 &  15.49 &  0.02 &  15.26 &  0.03\\
107 &  01 57 03.55 & $-$73 46 16.0 &  17.09 &  0.02 &  16.98 &  0.05 &  16.77 &  0.09\\
108 &  01 57 10.58 & $-$74 05 49.1 &  16.05 &  0.02 &  16.04 &  0.02 &  15.88 &  0.04\\
109 &  01 57 35.09 & $-$74 19 37.1 &  16.53 &  0.02 &  16.47 &  0.03 &  16.32 &  0.04\\
110 &  01 57 45.02 & $-$74 00 00.4 &  15.14 &  0.02 &  15.04 &  0.02 &  14.87 &  0.03\\
111 &  01 57 51.29 & $-$74 20 24.8 &  16.50 &  0.02 &  16.54 &  0.04 &  16.44 &  0.06\\
112 &  01 58 44.88 & $-$73 53 23.1 &  15.19 &  0.02 &  15.05 &  0.02 &  14.77 &  0.02\\
113 &  01 59 01.43 & $-$74 24 15.9 &  15.14 &  0.01 &  14.99 &  0.02 &  14.84 &  0.02\\
114 &  01 59 03.50 & $-$74 30 24.9 &  14.67 &  0.02 &  14.57 &  0.02 &  14.37 &  0.03\\
115 &  01 59 25.89 & $-$74 15 28.0 &  14.78 &  0.01 &  14.65 &  0.02 &  14.38 &  0.02\\
116 &  01 59 32.05 & $-$73 56 08.2 &  15.24 &  0.02 &  15.08 &  0.02 &  14.87 &  0.02\\
117 &  01 59 40.95 & $-$74 27 06.0 &  16.27 &  0.02 &  16.26 &  0.04 &  16.12 &  0.09\\
118 &  01 59 49.34 & $-$74 23 48.0 &  17.35 &  0.03 &  17.30 &  0.08 &  17.13 &  0.10\\
119 &  02 00 03.41 & $-$74 06 24.3 &  16.05 &  0.02 &  16.05 &  0.04 &  15.80 &  0.07\\
120 &  02 00 31.97 & $-$74 50 38.1 &  16.64 &  0.02 &  16.58 &  0.03 &  16.42 &  0.10\\
121 &  02 01 02.89 & $-$73 55 48.5 &  17.20 &  0.02 &  17.18 &  0.05 &  17.05 &  0.09\\
122 &  02 01 09.38 & $-$73 40 37.9 &  16.35 &  0.02 &  16.21 &  0.04 &  16.09 &  0.10\\
123 &  02 01 26.95 & $-$74 40 14.8 &  15.21 &  0.02 &  15.15 &  0.02 &  14.97 &  0.03\\
124 &  02 01 33.99 & $-$74 09 09.4 &  16.36 &  0.02 &  16.29 &  0.03 &  16.08 &  0.05\\
125 &  02 01 59.96 & $-$74 51 56.6 &  16.10 &  0.01 &  15.94 &  0.03 &  15.59 &  0.05\\
126 &  02 03 09.38 & $-$73 55 27.6 &  16.91 &  0.02 &  16.72 &  0.05 &  16.51 &  0.08\\
127 &  02 03 11.72 & $-$74 03 34.5 &  17.28 &  0.02 &  17.20 &  0.05 &  17.05 &  0.09\\
128 &  02 03 17.45 & $-$73 39 28.4 &  14.85 &  0.02 &  14.68 &  0.06 &  14.43 &  0.03\\
129 &  02 03 32.50 & $-$73 58 33.7 &  17.00 &  0.03 &  17.00 &  0.07 &  16.78 &  0.09\\
130 &  02 03 33.40 & $-$74 01 15.8 &  14.17 &  0.02 &  14.16 &  0.03 &  13.84 &  0.03\\
131 &  02 03 52.08 & $-$74 06 59.8 &  15.22 &  0.02 &  15.04 &  0.04 &  14.92 &  0.02\\
132 &  02 04 19.90 & $-$74 45 48.8 &  16.94 &  0.03 &  16.89 &  0.04 &  16.67 &  0.08\\
133 &  02 04 30.52 & $-$74 07 51.5 &  16.31 &  0.02 &  16.15 &  0.03 &  16.04 &  0.04\\
134 &  02 04 40.50 & $-$74 53 45.6 &  15.58 &  0.03 &  15.39 &  0.03 &  15.23 &  0.05\\
135 &  02 04 42.14 & $-$73 46 48.2 &  15.65 &  0.02 &  15.55 &  0.02 &  15.34 &  0.03\\
136 &  02 04 58.60 & $-$74 31 58.3 &  16.69 &  0.02 &  16.58 &  0.03 &  16.48 &  0.06\\
137 &  02 05 13.61 & $-$74 08 25.0 &  15.42 &  0.02 &  15.23 &  0.02 &  15.03 &  0.05\\
138 &  02 05 15.55 & $-$74 11 26.8 &  16.90 &  0.03 &  17.00 &  0.05 &  16.82 &  0.09\\
139 &  02 06 16.16 & $-$74 10 51.1 &  16.96 &  0.03 &  17.08 &  0.06 &  16.97 &  0.10\\
140 &  02 06 21.48 & $-$74 37 19.6 &  14.94 &  0.03 &  14.89 &  0.02 &  14.80 &  0.05\\
141 &  02 06 36.90 & $-$74 05 38.3 &  17.11 &  0.03 &  17.10 &  0.05 &  16.85 &  0.09\\
142 &  02 06 41.33 & $-$74 20 36.2 &  15.82 &  0.02 &  15.76 &  0.02 &  15.59 &  0.04\\
143 &  02 06 45.18 & $-$74 27 47.5 &  14.96 &  0.01 &  14.93 &  0.02 &  14.79 &  0.02\\
144 &  02 06 47.75 & $-$74 21 21.6 &  16.26 &  0.02 &  16.10 &  0.03 &  15.97 &  0.04\\
145 &  02 07 03.07 & $-$74 53 37.6 &  17.40 &  0.04 &  17.22 &  0.05 &  16.97 &  0.10\\
146 &  02 07 11.53 & $-$74 44 20.7 &  17.02 &  0.03 &  17.05 &  0.05 &  16.95 &  0.10\\
147 &  02 07 17.97 & $-$74 09 36.0 &  16.26 &  0.02 &  16.20 &  0.04 &  15.90 &  0.05\\
148 &  02 07 32.23 & $-$75 00 37.4 &  14.87 &  0.02 &  14.73 &  0.01 &  14.49 &  0.02\\
149 &  02 07 33.24 & $-$74 47 48.7 &  14.97 &  0.01 &  14.92 &  0.01 &  14.77 &  0.03\\
150 &  02 07 33.27 & $-$74 01 03.3 &  15.68 &  0.04 &  15.60 &  0.02 &  15.51 &  0.03\\
151 &  02 07 34.11 & $-$74 20 08.5 &  13.99 &  0.02 &  14.00 &  0.01 &  13.89 &  0.02\\
152 &  02 07 37.06 & $-$74 32 50.8 &  17.00 &  0.02 &  16.92 &  0.04 &  16.69 &  0.10\\
153 &  02 07 54.58 & $-$74 33 25.4 &  16.32 &  0.02 &  16.28 &  0.03 &  16.11 &  0.07\\
154 &  02 08 06.80 & $-$74 18 04.6 &  14.32 &  0.01 &  14.22 &  0.01 &  13.90 &  0.02\\
155 &  02 08 09.54 & $-$74 19 15.3 &  15.89 &  0.02 &  15.83 &  0.02 &  15.61 &  0.03\\
156 &  02 08 17.33 & $-$74 23 38.3 &  15.65 &  0.02 &  15.57 &  0.02 &  15.31 &  0.03\\
157 &  02 08 19.54 & $-$74 02 02.0 &  13.49 &  0.01 &  13.44 &  0.01 &  13.33 &  0.02\\
158 &  02 08 21.16 & $-$73 41 06.3 &  17.39 &  0.03 &  17.22 &  0.04 &  17.11 &  0.09\\
159 &  02 08 24.08 & $-$74 57 52.3 &  15.22 &  0.03 &  15.15 &  0.02 &  15.01 &  0.03\\
160 &  02 08 32.61 & $-$74 22 33.8 &  16.80 &  0.02 &  16.91 &  0.05 &  16.72 &  0.10\\
161 &  02 08 35.46 & $-$74 58 55.0 &  17.03 &  0.03 &  16.98 &  0.07 &  16.71 &  0.10\\
162 &  02 08 38.60 & $-$74 19 41.9 &  16.23 &  0.01 &  16.22 &  0.03 &  15.95 &  0.05\\
163 &  02 08 42.74 & $-$74 34 18.4 &  16.32 &  0.02 &  16.27 &  0.03 &  16.18 &  0.09\\
164 &  02 09 00.10 & $-$73 59 51.5 &  16.17 &  0.02 &  16.25 &  0.03 &  16.12 &  0.05\\
165 &  02 09 01.25 & $-$74 06 07.1 &  17.04 &  0.03 &  17.02 &  0.05 &  16.78 &  0.08\\
166 &  02 09 14.02 & $-$74 30 40.3 &  15.63 &  0.01 &  15.55 &  0.02 &  15.38 &  0.08\\
167 &  02 09 16.01 & $-$73 58 57.0 &  16.44 &  0.02 &  16.37 &  0.02 &  16.24 &  0.05\\
168 &  02 09 20.63 & $-$74 23 21.1 &  16.52 &  0.02 &  16.46 &  0.04 &  16.22 &  0.07\\
169 &  02 09 29.63 & $-$74 30 35.3 &  16.37 &  0.02 &  16.30 &  0.03 &  16.03 &  0.05\\
170 &  02 09 31.29 & $-$74 07 56.0 &  16.71 &  0.02 &  16.71 &  0.06 &  16.47 &  0.09\\
171 &  02 09 34.56 & $-$74 27 11.8 &  14.35 &  0.02 &  14.27 &  0.02 &  14.10 &  0.02\\
172 &  02 09 47.39 & $-$73 45 34.7 &  16.32 &  0.03 &  16.38 &  0.05 &  16.26 &  0.05\\
173 &  02 09 56.11 & $-$74 18 25.2 &  16.77 &  0.02 &  16.75 &  0.05 &  16.57 &  0.09\\
174 &  02 09 59.55 & $-$73 41 27.1 &  17.59 &  0.03 &  17.46 &  0.06 &  17.28 &  0.10\\
175 &  02 10 06.73 & $-$74 47 00.1 &  14.51 &  0.03 &  14.32 &  0.04 &  14.17 &  0.02\\
176 &  02 10 45.27 & $-$74 10 50.5 &  15.76 &  0.01 &  15.71 &  0.02 &  15.46 &  0.03\\
177 &  02 10 47.92 & $-$73 52 22.5 &  15.34 &  0.02 &  15.20 &  0.02 &  14.97 &  0.03\\
178 &  02 10 54.97 & $-$74 05 26.1 &  15.83 &  0.02 &  15.95 &  0.02 &  15.84 &  0.04\\
179 &  02 11 00.49 & $-$73 39 20.4 &  16.43 &  0.06 &  16.31 &  0.05 &  16.02 &  0.09\\
180 &  02 11 13.04 & $-$74 13 46.9 &  16.62 &  0.02 &  16.68 &  0.04 &  16.54 &  0.09\\
181 &  02 11 18.80 & $-$74 29 55.6 &  14.31 &  0.06 &  14.23 &  0.06 &  13.85 &  0.08\\
182 &  02 11 20.67 & $-$73 45 33.3 &  17.43 &  0.02 &  17.39 &  0.05 &  17.22 &  0.10\\
183 &  02 11 29.18 & $-$73 56 15.2 &  16.54 &  0.02 &  16.56 &  0.03 &  16.36 &  0.06\\
184 &  02 11 38.24 & $-$73 39 27.0 &  14.10 &  0.03 &  13.99 &  0.06 &  13.71 &  0.02\\
185 &  02 11 38.26 & $-$74 05 58.3 &  16.82 &  0.02 &  16.70 &  0.04 &  16.50 &  0.06\\
186 &  02 11 43.72 & $-$74 17 38.6 &  14.23 &  0.01 &  14.18 &  0.01 &  14.02 &  0.02\\
187 &  02 12 25.79 & $-$74 22 35.2 &  15.41 &  0.02 &  15.34 &  0.02 &  15.25 &  0.03\\
188 &  02 12 35.51 & $-$73 42 16.8 &  17.20 &  0.03 &  17.01 &  0.05 &  16.83 &  0.06\\
189 &  02 12 38.38 & $-$74 22 44.9 &  16.23 &  0.02 &  16.23 &  0.03 &  16.08 &  0.06\\
190 &  02 12 39.40 & $-$74 57 51.1 &  15.71 &  0.02 &  15.63 &  0.02 &  15.50 &  0.03\\
191 &  02 12 58.10 & $-$74 54 28.4 &  15.08 &  0.02 &  14.99 &  0.03 &  14.60 &  0.02\\
192 &  02 13 58.31 & $-$74 06 48.2 &  16.17 &  0.03 &  16.28 &  0.04 &  16.18 &  0.05\\
193 &  02 14 12.83 & $-$74 40 10.4 &  14.73 &  0.01 &  14.58 &  0.02 &  14.26 &  0.03\\
194 &  02 14 40.19 & $-$74 12 05.7 &  15.85 &  0.03 &  15.74 &  0.02 &  15.49 &  0.04\\
195 &  02 14 42.36 & $-$74 12 12.0 &  15.38 &  0.03 &  15.24 &  0.02 &  14.96 &  0.03\\
196 &  02 14 45.26 & $-$74 07 44.6 &  16.12 &  0.03 &  16.05 &  0.03 &  15.74 &  0.04\\
197 &  02 15 32.54 & $-$74 07 51.8 &  15.45 &  0.03 &  15.36 &  0.02 &  15.26 &  0.03\\
198 &  02 15 53.19 & $-$74 26 38.7 &  15.75 &  0.03 &  15.56 &  0.09 &  15.34 &  0.03\\
199 &  02 16 01.20 & $-$74 08 30.4 &  15.78 &  0.03 &  15.63 &  0.03 &  15.38 &  0.03\\
200 &  02 16 43.89 & $-$74 00 04.9 &  14.96 &  0.02 &  14.88 &  0.03 &  14.54 &  0.03\\
201 &  02 16 51.67 & $-$74 13 33.0 &  14.88 &  0.03 &  14.83 &  0.06 &  14.53 &  0.05\\
202 &  02 17 02.55 & $-$73 40 07.5 &  14.83 &  0.03 &  14.72 &  0.08 &  14.40 &  0.03\\
203 &  02 17 43.06 & $-$74 20 21.2 &  15.74 &  0.01 &  15.65 &  0.02 &  15.42 &  0.04\\
\enddata
\end{deluxetable}


\newpage

\end{document}